%% file: main.tex
\documentclass[10pt,conference]{IEEEtran}
\usepackage{cite}
\usepackage{amsmath,amssymb,amsfonts}
\usepackage{graphicx}
\usepackage{textcomp} % For trademark symbol
\usepackage{xcolor}
\usepackage{multicol} % Put two equations side by side
\usepackage{tikz}
\usepackage{circledsteps}
\usepackage[binary-units=true]{siunitx} % For GB formatting and scientific notation

\usepackage[pscoord]{eso-pic}
\usepackage[a4paper, total={184mm,239mm}]{geometry}
\def\BibTeX{{\rm B\kern-.05em{\sc i\kern-.025em b}\kern-.08em
    T\kern-.1667em\lower.7ex\hbox{E}\kern-.125emX}}

\usepackage[utf8]{inputenc}
\usepackage{listings}
\usepackage{xcolor}
\usepackage[xindy,style=index,numberedsection=false]{glossaries}
\usepackage{url}
\usepackage{multirow}
\usepackage[]{algpseudocode,algorithm}
\algtext*{EndWhile}% Remove "end while" text
\algtext*{EndIf}% Remove "end if" text
\algtext*{EndFor}% Remove "end if" text
\algtext*{EndFunction}% Remove "end if" text
\usepackage{textcomp}
\usepackage{cleveref}
\usepackage{balance}
\usepackage{graphicx}
\usepackage{subcaption}
\usepackage{booktabs}
\usepackage{makecell}
\usepackage{caption}
\usepackage{setspace}
\let\Algorithm\algorithm
\renewcommand\algorithm[1][]{\Algorithm[#1]\setstretch{1.0}}
\algnewcommand{\LineComment}[1]{\State \(\triangleright\) #1}

\crefname{algocf}{alg.}{alg.}
\Crefname{algocf}{Alg.}{Alg.}
\crefname{lstlisting}{listing}{listings}
\Crefname{lstlisting}{Listing}{Listings}
\crefname{algorithm}{alg.}{alg.}
\Crefname{algorithm}{Alg.}{Alg.}

\AtBeginDocument{} % Used in double-equations on one line;

\input{commands}

\input{acronyms}
    
\begin{document}

\title{Scaling up HBM Efficiency of Top-K SpMV for Approximate Embedding Similarity on FPGAs}

\author{
\IEEEauthorblockN{Alberto Parravicini\IEEEauthorrefmark{1}, Luca Giuseppe Cellamare\IEEEauthorrefmark{2}, Marco Siracusa\IEEEauthorrefmark{2} Marco D. Santambrogio\IEEEauthorrefmark{1}}
\IEEEauthorblockA{Politecnico di Milano, DEIB, Milan, Italy}
\IEEEauthorrefmark{1}{\{alberto.parravicini, marco.santambrogio\}}@polimi.it,
\IEEEauthorrefmark{2}{\{lucagiuseppe.cellamare, marco.siracusa\}}@mail.polimi.it
% \IEEEauthorrefmark{3}marco.siracusa@mail.polimi.it,
% \IEEEauthorrefmark{4}marco.santambrogio@polimi.it
% \vspace{20pt}
}

\maketitle

\begin{abstract}

% ****** Other version ******
%Top-K SpMV is a key component of similarity-search through numerical embeddings.
%Due to its sparse workload characteristics, Top-K SpMV does not generally perform efficiently on general-purpose NUMA systems implementing traditional caching strategies.
%Therefore, we propose a Top-K SpMV FPGA design that, leveraging reduced precision and a novel packet-wise CSR matrix compression, allows a custom data layout delivering high bandwidth efficiency.
%With HBM-based boards, we achieve 100x performance speedup over a multi-threaded CPU implementation and 2x speedup over a GPU with 20\% higher bandwidth, also with a 7.6x higher power-efficiency.
% ***************************

% 100 words abstract, required by DAC
Top-K SpMV is a key component of similarity-search on sparse embeddings. This sparse workload does not perform well on general-purpose NUMA systems that employ traditional caching strategies.
Instead, modern FPGA accelerator cards have a few tricks up their sleeve. We introduce a Top-K SpMV FPGA design that leverages reduced precision and a novel packet-wise CSR matrix compression, enabling custom data layouts and delivering bandwidth efficiency often unreachable even in architectures with higher peak bandwidth.
With HBM-based boards, we are 100x faster than a multi-threaded CPU implementation and 2x faster than a GPU with 20\% higher bandwidth, with 14.2x higher power-efficiency.

\end{abstract}


\glsresetall
\input{sections/intro.tex}
\input{sections/soa.tex}
\input{sections/problem.tex}

\input{sections/implementation.tex}
\input{sections/experimental_results.tex}
\input{sections/conclusion.tex}

\bibliographystyle{IEEEtran}
% Balance columns, put a column break after reference 22
% \IEEEtriggeratref{23}

\bibliography{references}

\end{document}

%% file: commands.tex
\newcommand{\ra}[1]{\renewcommand{\arraystretch}{#1}}
\newcommand{\pluseq}{\mathrel{+}=}

%% file: acronyms.tex
\newacronym{nlp}{NLP}{Natural Language Processing}
\newacronym{ppr}{PPR}{Personalized PageRank}
\newacronym{spmv}{SpMV}{Sparse matrix-vector multiplication}
\newacronym{topkspmv}{Top-K SpMV}{Top-K Sparse matrix-vector multiplication}
\newacronym{coo}{COO}{Coordinate}
\newacronym{dsl}{DSL}{Domain-Specific Language}
\newacronym{csc}{CSC}{Compressed Sparse Column}
\newacronym{csr}{CSR}{Compressed Sparse Row}
\newacronym{raw}{RAW}{Read-After-Write}
\newacronym{ndcg}{NDCG}{Normalized Discounted Cumulative Gain}
\newacronym{ir}{IR}{Information Retrieval}
\newacronym{dcg}{DCG}{Discounted Cumulative Gain}
\newacronym{er}{ER}{Entity Resolution}
\newacronym{ii}{II}{Initiation Interval}
\newacronym{gpu}{GPU}{GPU}
\newacronym{fpga}{FPGA}{FPGA}
\newacronym{colamd}{COLAMD}{ColumnApproximate Minimum Degree algorithm}
\newacronym{cpu}{CPU}{Central Processing Unit}
\newacronym{hbm}{HBM}{High Bandwidth Memory}
\newacronym{bscsr}{BS-CSR}{Block-Streaming CSR}
\newacronym{glove}{GloVe}{Global Vectors for Word Representation}

%% file: sections/intro.tex
\section{Introduction}\label{sec:intro}

\gls{ir} and recommender system process an always-increasing amount of data, often with strong real-time constraint, to suggest products, movies, news articles to billions of users.
\gls{topkspmv} is a key building block of high-performance similarity-search applications found in recommender systems that store items or documents as sparse \textit{embeddings}, short numerical representations usually obtained through a neural network \cite{sparsedottopn, aumller2018annbenchmarks}. Sparse embeddings guarantee low memory footprint and reduced execution time, often without decreasing accuracy, as they capture important information while filtering out noise \cite{berend-2017-sparse}.
\gls{topkspmv} computes the highest K values of the product between a sparse matrix (a matrix where only non-zero entries are stored) and a dense vector. In our case, it matches an input embedding against a collection of sparse embeddings and finds the K most similar ones (\Cref{fig:topk-spmv}). 

% The computation of \gls{topkspmv} is extremely memory intensive since including indirect, fully-random memory accesses. For highly sparse datasets, this behavior substantially challenges cache policies of general-purpose architectures, delivering poor performance \cite{paperBerkeley}.

The computation of \gls{topkspmv} is extremely memory intensive and presents sequential, indirect, and fully random memory accesses, making it unsuitable for general-purpose architectures with traditional caching policies. Improving its performance through a custom hardware design demands a theoretically sound approach to optimize the data-access strategy.
FPGAs can fully leverage application-specific data representations thanks to reduced-precision arithmetic and a customizable memory subsystem. In agreement with the Roofline methodology \cite{Siracusa2020RooflineFPGA}, this approach is key to improving operational intensity and performance in memory-starved computations.
Exact numerical accuracy for large values of K is not critical, as long as the most similar embeddings are retrieved:
compared to CPUs or GPUs, FPGAs can leverage optimized arbitrary-precision arithmetic and offer fine-grained control over the desired target accuracy, providing better performance, lower resource utilization, and lower power consumption.
\glsunset{hbm}
To fully utilize off-chip bandwidth, we devise an approximation scheme based on matrix partitioning that enables a flexible multi-core design with a single \gls{hbm} channel per core and URAM for caching high-reuse or random access data (\cref{sec:approximation}). 
On an Alveo U280 accelerator card, our design accesses 32 HBM2 pseudo-channels with a total bandwidth of 460 GB/s.
Then, we maximize bandwidth efficiency through a novel streaming sparse matrix format that leverages coalesced HBM accesses and reduced numerical precision, and is oblivious to the matrix non-zero entries distribution (\cref{sec:bscsr}).
We propose a novel multi-core streaming FPGA design to approximate the computation of \gls{topkspmv}\footnote{Code is available at \url{github.com/AlbertoParravicini/approximate-spmv-topk}}
and present the following contributions:

\begin{figure}[t]
    \centering
    \includegraphics[width=0.7\columnwidth]{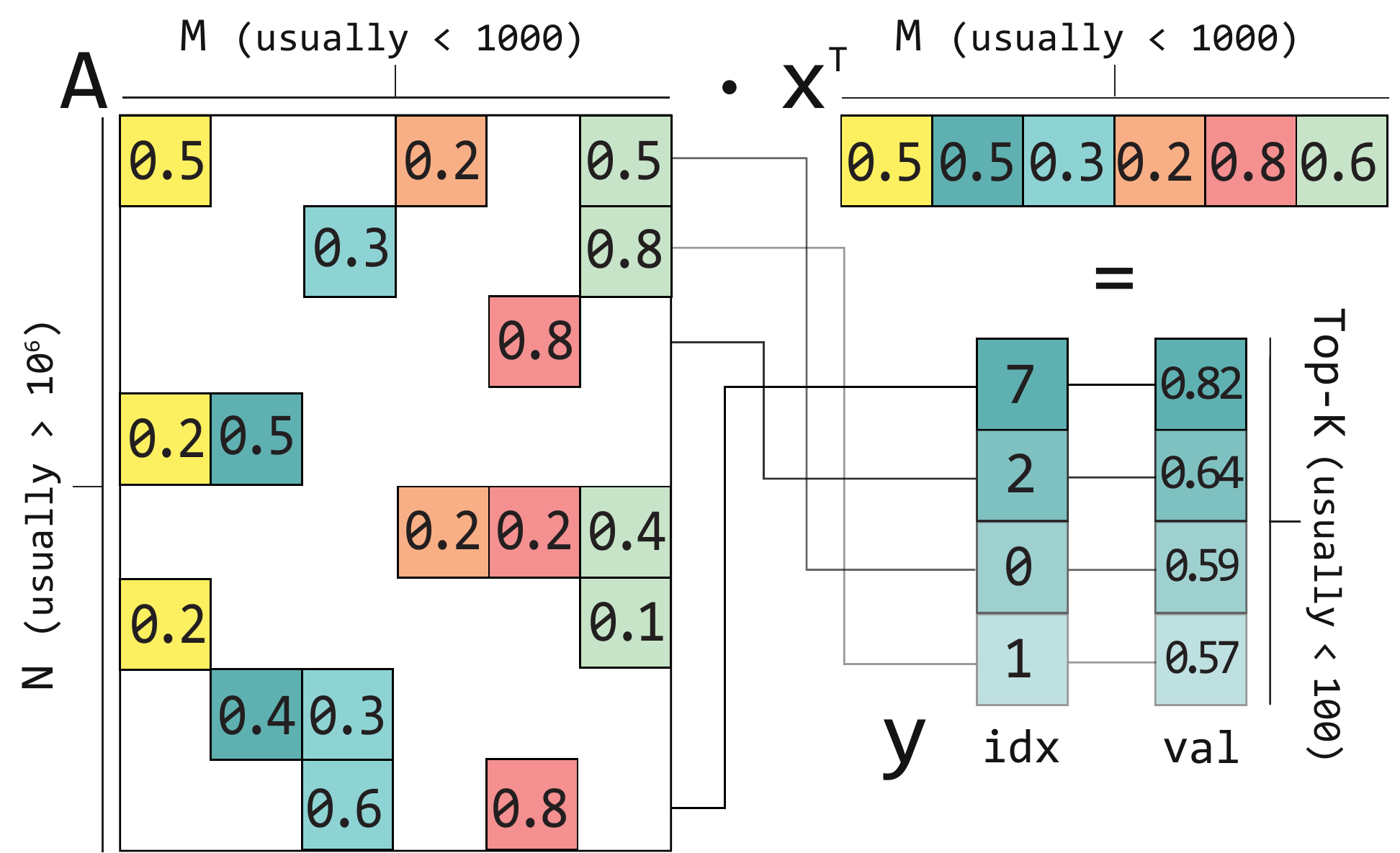}\\
    \caption{\gls{topkspmv} between a sparse matrix $\mathbf{A}$ (in our case, a collection of sparse embeddings) and a dense vector $\mathbf{x}$ (a dense embedding), with notation as in \Cref{sec:problem}.}
    \label{fig:topk-spmv}
\end{figure}

\begin{itemize}
    \item A multi-core \gls{topkspmv} FPGA design that leverages a novel compressed matrix representation and algorithmic approximations (\cref{sec:problem}) and  to effectively exploit the available \gls{hbm} bandwidth (\Cref{sec:architecture}).
    \item A performance evaluation of our FPGA design, showing 100x faster execution versus a state-of-the-art CPU implementation. We are also 2x faster than a GPU with 20\% higher bandwidth, with 14.2x higher power-efficiency and no significant loss of Top-K precision (\Cref{sec:experimental_results}).
\end{itemize}

%% file: sections/soa.tex
\section{Related Work}

% The optimization of \gls{spmv} is a challenging task in high-performance computing, due to the numerous factors degrading its performance \cite{grossman2016survey}, like for example the imbalances in terms of non-zero values per rows, storage format and the memory bandwidth of the hardware platform it executes on.

To the best of our knowledge, no prior work optimizes the computation of \gls{topkspmv} on FPGA or GPU, although existing research covers approximation techniques on FPGA for dense matrix multiplications and deep learning  \cite{doi:10.1007/s11265-020-01582-7, 8701880}.
On CPUs, \texttt{sparse\_dot\_topn} is a C++ multi-threaded implementation that employs the \gls{csr} format and is specialized for sparse embeddings and documents similarity \cite{sparsedottopn}. That said, the CPU performance in this computation is inherently held back by their limited memory bandwidth and the inability to consistently perform quick random accesses, as there are no guarantees that requested values have not been evicted from cache.
% Another approach to approximate similarity-search is FAISS: it retrieves the Top-K dense vectors by approximating the k-nearest neighbors on GPU or FPGA \cite{ faissfpga}.
\glsunset{spmv}
Although no existing work specifically optimizes \gls{topkspmv} on GPU, there exist multiple highly-optimized implementations of \gls{spmv} \cite{naumov2010cusparse,liu2015lightspmv,steinberger2016naive,liu2015csr5}. While modern GPUs provide higher memory bandwidth than FPGA accelerator cards (from 549 GB/s of the Nvidia P100 to the 1555 GB/s of the Nvidia A100), existing \gls{spmv} implementations are often unable to utilize the available bandwidth fully \cite{nguyen2020FPGAPotential}. Moreover, optimizations for reduced precision data-types are currently limited to half-precision floating-point, with no support for reduced-precision fixed-point arithmetic   \cite{abdelfattah2020survey,haidar2018harnessing}
% limits their maximum operational intensity and precise control over numerical accuracy (\Cref{sec:roofline}),
although recent work has explored heuristics to mix single and double precision floating-point arithmetic \cite{ahmad2019data}.
% \cite{p100,a100}

Existing research on \gls{spmv} optimization on FPGA focuses on sparse matrix compression \cite{grigoras2015accelerating}, moving the bottleneck from memory accesses to the input data decompression. Umuroglu et al. \cite{umuroglu2015vector} maximize the time in which values are kept in a fast local cache hierarchy.
The efficient use of \gls{hbm} on FPGA also has received the academic community's attention. Sadi et al. \cite{10.1145/3352460.3358330} propose an \gls{spmv} FPGA implementation that achieves significant speedup leveraging \gls{hbm} and a data-compression scheme to reduce off-chip traffic.
% It has also been shown how \gls{hbm} can provide an order of magnitude speedup over CPU on a variety of data-intensive workloads \cite{kara2020high}.
% Qui piazzaci un paio di self-cit se ti pigliano
However, directly employing off-the-shelf \gls{spmv} FPGA designs (e.g. \cite{parravicini2021reduced}) is undesirable: it would waste memory to store for the entire output vector (which can contain millions of values, \Cref{sec:problem}) and demand unnecessary computation to sort the output afterward. 

% BUTTO TUTTA LA ROBA EXTRA IN FONDO

% Numerous algorithmic optimizations for \gls{spmv} have been proposed,
% toward the goal of minimizing cache misses. \gls{colamd} proposes a column reordering to enable efficient use of memory hierarchy.
% Fowers et al. \cite{6861585} introduce a sparse matrix encoding to maximize bandwidth utilization.
%State-of-the-art \gls{hbm} has shown promising results in alleviating the memory accesses bottleneck of \gls{spmv} \cite{10.1145/3352460.3358330}, thanks 
% Moreover, reduced-precision arithmetic has not been deeply studied in the context of \gls{spmv} algorithms, but encouraging results have been obtained in the field of numerical analysis \cite{anson2012optimising, sun2008high} and deep-learning \cite{liang2018fp, wang2019deep, nagy2003configurable}.

%% file: sections/problem.tex
\section{Theoretical Contributions}\label{sec:problem}

\begin{figure}[t]
    \centering
    \includegraphics[width=0.8\columnwidth]{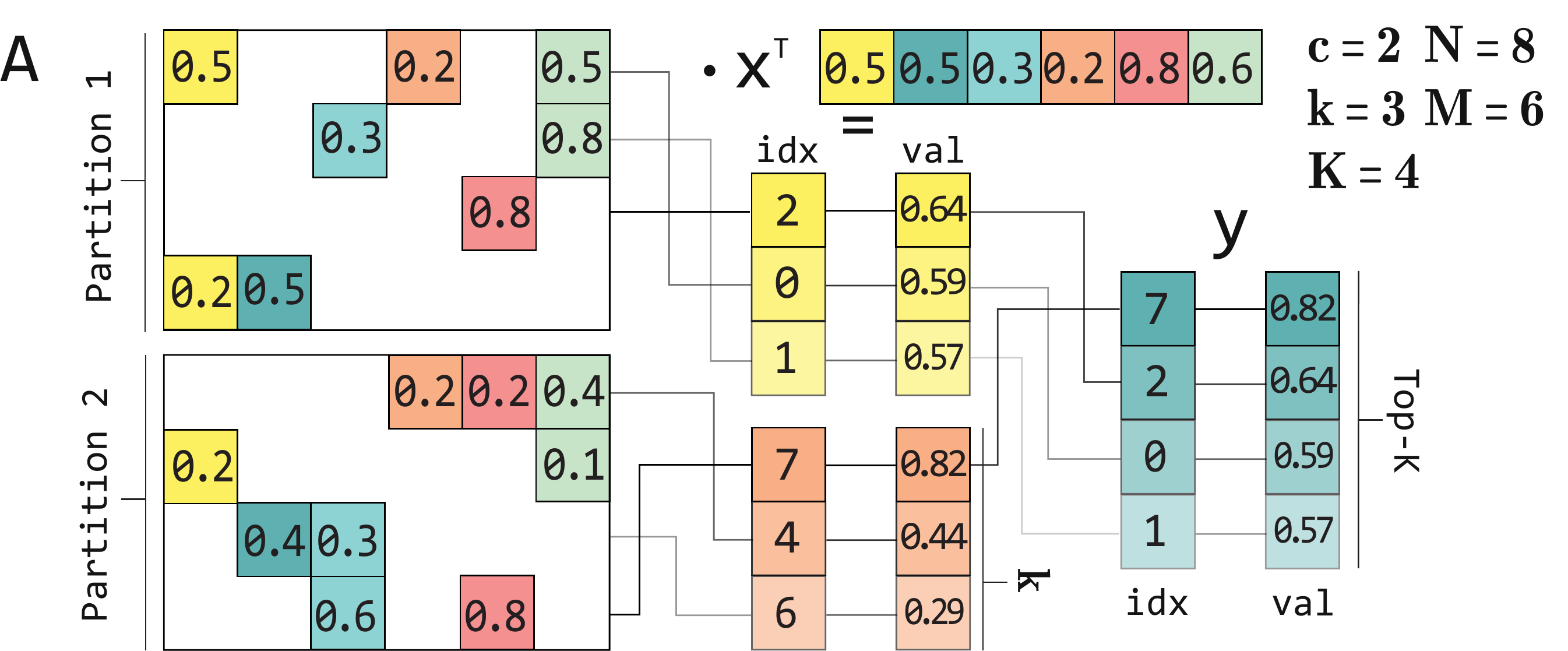}\\
    \caption{Simplified approximation scheme for \gls{topkspmv}. No errors occur if all partitions contain less than $\mathbf{k}$ Top-K values.}
    \label{fig:topk-spmv-approx}
\end{figure}

\input{tables/simulated_errors}

Two key theoretical contributions enable our FPGA hardware design. First, we introduce a partitioning scheme that splits and approximates \gls{topkspmv} on independent cores, giving access to the entire HBM bandwidth (\Cref{sec:approximation}). Then, we present \gls{bscsr}, a sparse matrix format optimized for streaming coalesced memory transactions and reduced-precision arithmetic (\Cref{sec:bscsr}).

Given a sparse matrix $\mathbf{A}$ of size $N \times M$, and a dense vector $\mathbf{x}$ of size $M \times 1$, the result $\mathbf{y}$ of the \gls{spmv} $\mathbf{y} = \mathbf{A}\mathbf{x}$ is a vector $N \times 1$. Top-K \gls{spmv} does not track the entire $\mathbf{y}$, but only its highest K values (and the corresponding indices). If $\mathbf{A}$ and $\mathbf{x}$ are $L2$-normalized embeddings, Top-K \gls{spmv} retrieves the rows of $\mathbf{A}$ with highest cosine similarity to $\mathbf{x}$.
In our task, $N$ is the size of the embedding collection (usually millions of entries), while $M$, the dense embedding size, has only a few hundred values \cite{yin2018dimensionality}. $\mathbf{x}$ is stored in URAM to perform random accesses in a single clock cycle, which is infeasible with an unconstrained $M$.
Similarly, we track just the Top-K entries of $\mathbf{y}$ on a LUT scratchpad instead of performing data-dependent write-backs on HBM and sharing bandwidth with read-channels, which would undermine the maximum memory throughput.
$\mathbf{x}$ is dense as many sparsification algorithms operate on dense matrices (e.g. batches of $\mathbf{A}$) 
and cannot efficiently sparsify a single vector \cite{tropp2007signal,mairal2009online}.
% aharon2006k K-SVD

\subsection{Top-K SPMV Approximation}\label{sec:approximation}

In Top-K similarity-search applications, it is often acceptable to trade maximum accuracy on large values of $K$ for lower execution time and power consumption.
Instead of computing the total Top-K values on the matrix $\textbf{A}$, we partition it in $\mathbf{c}$ sub-matrices with $N / \mathbf{c}$ rows each. Each sub-matrix is processed by an independent FPGA core, which computes the top $\mathbf{k} < K$ results for the partition (with $\mathbf{k}\cdot\mathbf{c} \geq K$). We obtain $\mathbf{k}\cdot\mathbf{c}$ results, which represents an approximation of the original Top-K (\Cref{fig:topk-spmv-approx}).
As $\mathbf{c}$ increases, so does the approximation accuracy. As we always retrieve the top $\mathbf{k}$ values, the approximation does not affect the best-ranked rows.
\Cref{eq:approximation} presents a closed-form expression of the expected value of correctly retrieved Top-K values (i.e. the \textit{precision}); intuitively, \Cref{eq:approximation} compares the number of permutations with more than $\mathbf{k}$ Top-K values in a partition with the total number of Top-K permutations.
We estimate $\mathbb{E}[P]$ through a Monte Carlo simulation for different $K$ and number of partitions, with $K$ from 8 to 100, common thresholds in \gls{ir} \cite{dacrema2019we}. Having at least 16 partitions guarantees a minimal loss of precision, even for large matrices.
\begin{equation}\label{eq:approximation}
    \mathbb{E}[P] \approx \frac{1}{K - 1} \sum_{K_i = 1}^{K}{1 - \frac{\mathbf{c} \cdot \sum_{\mathbf{k}_i = \mathbf{k} + 1}^{min(K_i, \lfloor N/\mathbf{c} \rfloor)}{\binom{\lfloor N/\mathbf{c} \rfloor}{\mathbf{k}_i}}}{\binom{N}{K_i}}}
\end{equation}

\subsection{The Block-Streaming CSR Matrix Layout}\label{sec:bscsr}

% The goal of this work is to present an optimized FPGA implementation of Top-K \gls{spmv}, a linear-algebra primitive that retrieves the highest $K$ values of the product between a sparse matrix and a dense vector.

% Top-K \gls{spmv} can compute the documents in a corpus (a collection of documents) that are the most similar to a document, or suggest movies that might interest a user who watched a certain film.

% The resulting vector yi
% can be obtained by multiplying matrix A by vector x . Each ith element of vector y is the result of cumulated multiplications of non-zero values inside ith row of matrix A and corresponding vector x values as shown in \cref{eq:ith result of spmv}
% \begin{equation}\label{eq:ith result of spmv}
%     \mathbf{y_i} = 
%     \sum_{j=0}^{N-1}\mathbf{A_{ij}}
%     \mathbf{x_j}, \quad 0 \le i \le {M-1}
% \end{equation}

The matrix $\mathbf{A}$ is stored in a \textit{sparse format} to save only non-zero values and lower its memory footprint.
Different storage techniques are possible, depending on the non-zero elements distribution and the desired type of access patterns.

The common \gls{csr} format \cite{yang2018design} is unsuitable for fully-pipelined streaming FPGA designs as it contains data dependencies to access matrix values, and hiding the memory controller latency is not trivial, given the lack of a built-in hardware prefetcher.
Instead, the \gls{coo} layout uses three equally sized arrays to store, for each matrix entry, its two coordinates and the entry itself's value. 
\gls{coo} allows streaming iterations with burst memory transactions on non-zero matrix entries to saturate bandwidth, as the architecture does not have to perform data-dependent memory accesses determined by the number of non-zero values of each row, as in the \gls{csr} format. 
\gls{coo} allows coalesced memory transactions of multiple non-zero entries, but requires redundant storage of coordinate values, limiting the overall operational intensity. 

% The \gls{coo} format simplifies array partitioning, as computation on each non-zero value can be done independently from others.
% This format enables higher throughput and better hardware scalability up to the available FPGA resources.

\begin{figure}[t]
    \centering
    \includegraphics[width=\columnwidth]{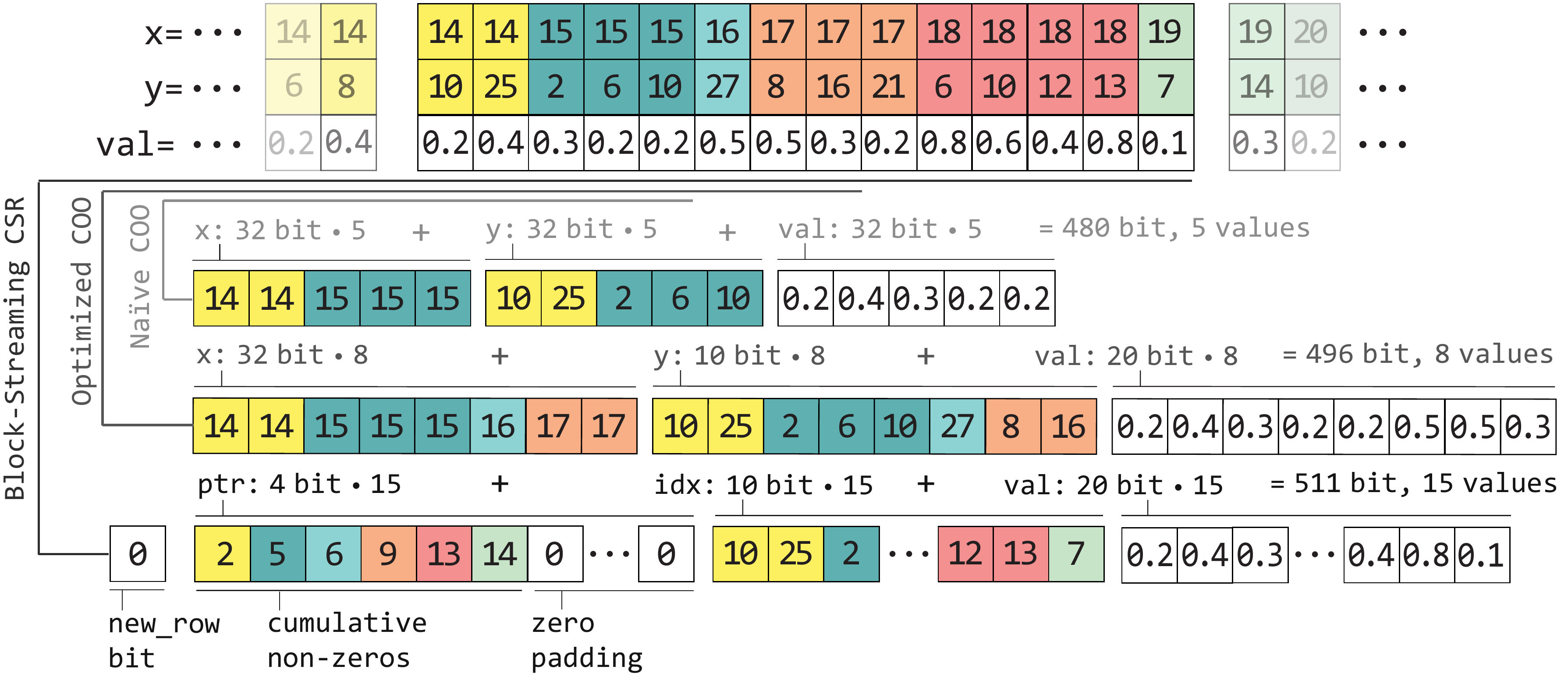}\\
    \caption{Na\"ive \gls{coo} allows only 5 non-zero entries in a 512-bits packet. Using fewer bits for \texttt{y} and \texttt{val} (assuming \texttt{y} $< 1024$ and \texttt{val} stored as 20-bits fixed-point) permits only 3 additional non-zeros per packet. \gls{bscsr} drastically reduces the space required for \texttt{ptr}, enabling 15 non-zeros per packet.}
    \label{fig:bscsr}
\end{figure}

We propose \gls{bscsr}, a new sparse matrix storage format that combines the low memory footprint of \gls{csr} with the streaming properties of \gls{coo}, and is enabled by reduced precision data-types.
As the architecture of the Alveo U280 HBM memory controllers incentivizes memory transactions of data packets (or blocks) between 256 and 512 bits \cite{wang2020shuhai}, we build each packet in \gls{bscsr} as an independent \gls{csr} partition (\Cref{fig:bscsr}): values of \texttt{idx} and \texttt{val} are identical to standard \gls{csr}, using less than 32 bits for each value if possible (in our case, 10 bits are enough for \texttt{idx} entries). The \texttt{ptr} vector tracks the cumulative non-zero entries with respect to the packet itself: each entry of \texttt{ptr} is just $\lfloor log_2(B)\rfloor$ bits for a packet with $B$ non-zero entries (4 bits for a packet $B\!=\!15$), instead of 32 bits like in a na\"ive \gls{csr} or \gls{coo} (as the number of rows is not bound). The $new\_row$ bit tracks if the first row of a packet continues the last row of the previous packet. Missing rows are handled with placeholder $0$ values, although rows are never fully empty in our application domain. \gls{bscsr} does not provide an explicit reference to row indices: this is not a limitation in streaming algorithms such as \gls{spmv} and its variations, as we process the entire matrix sequentially and track the current rows by counting non-zero entries in each \texttt{ptr} packet, plus the $new\_row$ bit. We can fit 2 to 3 times as many non-zero entries in each 512-bits packet and improve the operational intensity by the same amount. Although significant work has been made to optimize \gls{spmv} on matrices with irregular row density distribution  \cite{10.1145/3352460.3358330}, we avoid the problem entirely through a fully-streaming data format that is oblivious to the number of non-zero matrix entries per row.
In spite of its simplicity, \gls{bscsr} can be a life-saver in streaming memory-bound computations (\Cref{sec:roofline}).

%% file: tables/simulated_errors.tex
\renewcommand\theadalign{tc}
\renewcommand\theadfont{\bfseries}
\setlength\tabcolsep{5pt}

\begin{table}
\centering
\ra{1.2}
    \caption{Estimated precision of Top-K indices for increasing number of partitions. We average results of 1000 tests.}
    \resizebox{\columnwidth}{!}{
	\begin{tabular}{@{}ccccccccc@{}}
		\toprule
		\multirow{2}{*}{\thead{Number of\\matrix rows}} &	\multirow{2}{*}{\thead{Number of\\partitions}} &  
		\phantom{abc} & \multicolumn{6}{c}{\thead{Top-K Value, \textnormal{with} $\mathbf{k}=8$}} \\
		\cmidrule{4-9}
		 &&& \thead{8} & \thead{16}  & \thead{32}  & \thead{50}  & \thead{75}  & \thead{100} \\
		\midrule
		
		\multirow{3}{*}{$N = 10^6$} & $\mathbf{c}=16$ && $1$ & $1$ & $0.999$ & $0.998$ & $0.983$ & $0.942$ \\ 
		                        %& $21$ && $1$ & $1$ & $1$ & $0.999$ & $0.996$ & $0.982$ \\
	                           % & $26$ && $1$ & $1$ & $1$ & $0.999$ & $0.998$ & $0.993$ \\
                		        & $\mathbf{c}=28$ && $1$ & $1$ & $1$ & $0.999$ & $0.999$ & $0.996$ \\
                		        & $\mathbf{c}=32$ && $1$ & $1$ & $1$ & $0.999$ & $0.999$ & $0.997$ \\
    	\midrule
    	\multirow{3}{*}{$N = 10^7$} & $\mathbf{c}=16$ && $1$ & $1$ & $1$ & $0.999$ & $0.986$ & $0.947$ \\
    	                        %& $21$ && $1$ & $1$ & $1$ & $0.999$ & $0.996$ & $0.982$ \\
                	           % & $26$ && $1$ & $1$ & $1$ & $0.999$ & $0.999$ & $0.994$ \\
                		        & $\mathbf{c}=28$ && $1$ & $1$ & $1$ & $0.999$ & $0.999$ & $0.995$ \\
                		        & $\mathbf{c}=32$ && $1$ & $1$ & $1$ & $0.999$ & $0.998$ & $0.998$ \\
		\bottomrule
	\end{tabular}
    }
    \label{tab:simulated_errors}
\end{table}

%% file: sections/implementation.tex
%This section details the proposed top-k \gls{spmv} FPGA accelerator. Its main building block is discussed with the optimizations made to leverage information about documents in our intended use-case, and the integration with a CPU-based host system.

\section{The Proposed FPGA Hardware Design}\label{sec:architecture}

Our Top-K \gls{spmv} FPGA design offers multiple low-profile cores that operate independently. Each core processes a partition of the input matrix, as explained in \Cref{sec:approximation}. 
To guarantee maximum flexibility in terms of the number of cores placed on the FPGA, each core uses a single \gls{hbm} channel to read the input matrix and store its output at the very end of the computation. As cores read 512 bits per clock cycle (at 225 MHz) from their respective \gls{hbm} memory port and memory transactions happen in continuous maximum length AXI4 bursts (256 beats), our FPGA design can theoretically operate at the maximum bandwidth offered by \gls{hbm}, by coalescing transactions with \gls{bscsr} packets containing multiple non-zero entries, and without expensive distributed memory controllers. Each core reads $\mathbf{A}$ in packets of size $B$. $B$ ranges from 7 to 15, depending on the desired level of numerical precision and the embedding size: at worst we use 32 bits for \texttt{idx} and \texttt{val}, but realistic size bounds (e.g. \texttt{idx} $< 1024$) allow much greater coalescing and operational intensity (\Cref{fig:roofline}).

\begin{figure}[t]
    \centering
    \includegraphics[width=\columnwidth]{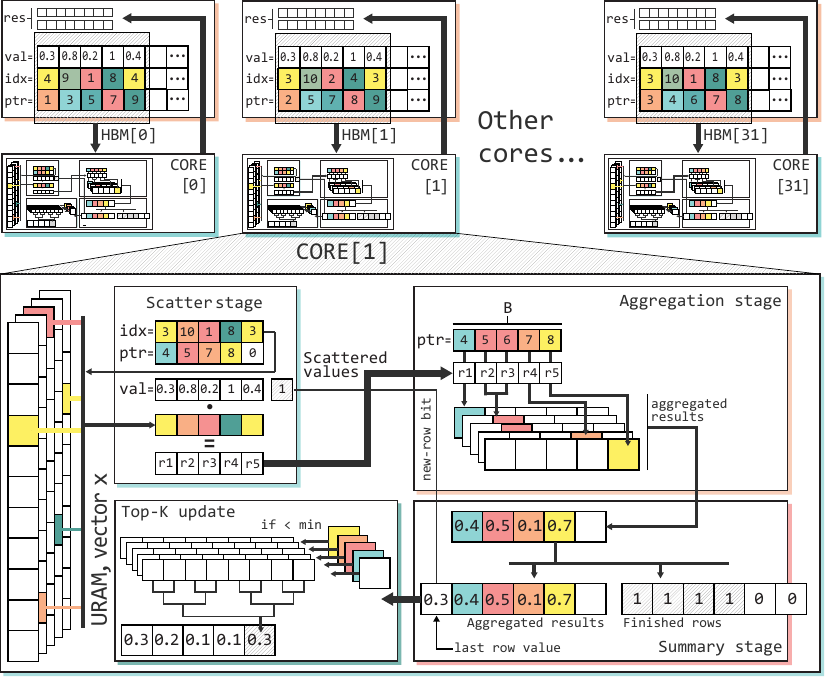}\\
    \caption{Block diagram of our multi-core Top-K SpMV, simplified for \gls{bscsr} with 5 non-zero \texttt{y} and \texttt{val} entries per packet.}
    \label{fig:architecture}
\end{figure}

\subsection{Leveraging URAM for Fast Random Access}\label{sec:uram}
The input vector $\mathbf{x}$ is stored in URAM.
Our core performs $B$ random accesses per cycle on $\mathbf{x}$. As each URAM bank only has 2 read ports, we replicate $\mathbf{x}$ $\lceil B / 2 \rceil$ times to allow all random accesses. This replication does not constitute a limitation of our design: $\mathbf{x}$ represents a dense embedding whose size does not go beyond a few hundred values in real applications. With our implementation, $\mathbf{x}$ can have size up to 80000 (assuming a worst-case scenario with 32 bits values, 32 cores, and 8 replicas of $\mathbf{x}$ per core), given a URAM size of around 90MB.

\subsection{Top-K SpMV Algorithm Design}\label{sec:algorithm}
The main Top-K \gls{spmv} algorithm (\cref{alg:spmv}) is implemented as a 4-stage pipelined data-flow computation, decoupling memory transfer and computation. Each stage processes the input matrix as packets of $B$ elements. The first stage retrieves $B$ entries from $\mathbf{A}$ per cycle from HBM and URAM and computes point-wise products. Values corresponding to the same matrix row are aggregated in the second stage. The third stage performs book-keeping for rows that span more than one packet and finds which rows have been fully processed in the current packet. The final stage updates the current Top-K values for each complete row if the row output value is in the Top-K.
Our data-flow design completely hides the Top-K update cost, which would otherwise be an expensive part of the computation.

% We store 5 different buffers of size 8 and update them independently. That is because a single packet can have > 1 finished row, and we need to update the top-k vector within a clock cycle to guarantee pipeline with ii=1. We compute 5*8 top k per core from a practical perspective, although most of the extra k-1 * 5 values will be empty as it is unlikely that many rows finish in a single packet.
Each core retrieves the top $\mathbf{k} = 8$ values of each matrix partition. Higher $\mathbf{k}$ results in lower clock speed due to RAW data-dependencies in the \texttt{argmin} computation, while lower $\mathbf{k}$ decreases accuracy.
% To avoid consumption of unnecessary logic, we can impose a limit on how many rows are entirely contained in a packet $B$: it is statistically unlikely that more than 
It is unlikely that many rows are entirely contained in a single packet $B$: we avoid consumption of unnecessary logic by tracking results for at most $\mathbf{r}$ rows per packet. In our experiments, using $B/4 < \mathbf{r} < B/2$ provided resource savings up to 50\% with no accuracy loss.
% The low clock frequency in our designs is caused by cores being placed in reconfigurable regions of the FPGA that do not have direct access to \gls{hbm}, requiring expensive routing logic. Having more cores at a lower frequency still provides better performance than fewer cores at a higher frequency.

\subsection{Lower Precision, More Cores, Better Performance}
% Values of $\mathbf{A}$ and $\mathbf{x}$ are encoded with reduced-precision fixed-point values: reduced-precision drastically lowers the amount of logic required by each core, allowing us to have 32 cores when using 20 bits fixed-point values, compared to the 24 cores of a 32-bits fixed-point design or the 21 of a 32-bits floating-point design. \Cref{tab:resources} summarizes the designs in our evaluation. 

Our hardware design processes $\mathbf{c}\cdot\!B$ non-zeros per clock cycle.
Maximizing this performance equation is not trivial, as increasing $B$ increments FPGA resource consumption and possibly prevents the placement of 32 cores, a requirement to achieve full HBM bandwidth utilization.
Indeed, $\mathbf{c}$ is a non-linear function of $\mathbf{k}$, $\mathbf{r}$ and $B$, with $B$ depending from $M$ and from the number of bits $V$ used for entries of \texttt{val}, as in $B\cdot (\lceil log_2{B}\rceil + \lceil log_2{M} \rceil + V) + 1\!=\!512$. $M$ can be safely assumed in the order of thousands. Choosing $V$, $\mathbf{k}$ and $\mathbf{r}$ requires a performance-accuracy trade-off: based on theoretical bounds in \Cref{tab:simulated_errors} and the experiments in \Cref{sec:accuracy}, we observe that $\mathbf{k} = 8$ and $V = 20$ are sufficient to provide great accuracy. Combined with $\mathbf{r}$ between 4 and 8 we guarantee a 32-cores design that can exploit all HBM channels without significant accuracy loss.
% Only with floating-point arithmetic, it was not possible to use 32 cores due to routing constraints.
\Cref{tab:resources} summarizes characteristics and resource usage of the designs in our evaluation. 

\input{code/spmv.tex}

% \subsection{Host Integration}\label{sec:host}
% The two proposed architectures follow both a standard host-accelerator model in which the \textit{host} (a traditional server) communicates with the \textit{accelerator} (an FPGA) over a PCIe interface.
% The host loads the matrix and pre-processes it, i.e. it computes the $val$ vector (the numerical entries of the matrix $\mathbf{X}$).
% These steps are done once at the start and do not require to be performed for each subsequent computation.
% Managing the FPGA execution from the host machine and handling the data transfers are done using the OpenCL API 1.2.
% The accelerator is implemented with Xilinx Vivado HLS and Xilinx SDx\textsuperscript{TM} 2018.3. Re-synthesizing the architecture is required if we change the fixed-point precision, the value of $\kappa$, or the maximum number of vertices stored on URAM, but not for different input matrices.

%% file: code/spmv.tex
\algrenewcommand\algorithmicindent{1.2em}%
\begin{algorithm}[t]
  \caption{Approximate \gls{bscsr} Top-K SpMV}\label{alg:spmv}
  \begin{algorithmic}[1]
    \Function{Top-K-SpMV}{$bscsr\_matrix, \mathbf{vec}$}
      \State $x_{s\_old} \gets 0;\ last\_packet\_output \gets 0$
      \For{$i \gets 0..\text{NNZ} / B$} \Comment{For each packet of the matrix}
        \LineComment{\textbf{1. Process \gls{bscsr} in packets of size $B$}}
         \For{$j \gets 0..B$} \Comment{All loops of size $B$ are Unrolled}
            \State $bscsr\_packet \gets bscsr\_matrix[i]$
            \State $x_{loc}[j] \gets bscsr\_packet.x[j]$
            \State $val_{loc}[j] \gets bscsr\_packet.val[j]$
            % \State $vec_{loc}[j] \gets $
            \State $res_{tmp}[j] \gets val_{loc}[j] \cdot \mathbf{vec}[j][bscsr\_packet.y[j]]$
        \EndFor
        
        \LineComment{\textbf{2. Aggregate partial $res_{loc}$ values}}
        % \For{$b1 \gets 0..B$} \Comment{Unrolled loops}
        %   \For{$b2 \gets 0..B$}
        %         \State $res_{agg}[x_{loc}[0]\ \%\ B + b1] \pluseq$\par
        %         \hskip\algorithmicindent $res_{tmp}[b2] \cdot ((x_{loc}[0] + b1) == x_{loc}[b2])$
        %     \EndFor
        % \EndFor
        \For{$b1 \gets 0..B$} \Comment{$x_{loc}[-1] = 0$ by convention}
           \State $row_{curr} \pluseq (x_{loc}[b1] \neq x_{loc}[b1-1])$
           \For{$b2 \gets x_{loc}[b1-1]..x_{loc}[b1]$} 
                \State $res_{agg}[b1] \pluseq res_{tmp}[b2]$
            \EndFor
        \EndFor
        \State $row_{curr} \pluseq bscsr\_packet.new\_row - 1$
            
        \LineComment{\textbf{3. Check if first value was split among packets}}
        \For{$j \gets 1..B$} \Comment{Find finished rows}
            \State $rows_{fin}[j] \gets x_{loc}[j - 1] \neq 0$
        \EndFor
        \If{$bscsr\_packet.new\_row$} 
            \State $res_{agg}[0] \gets last\_packet\_output$
            \State $rows_{fin}[0] \gets true$
        \Else  \Comment{This packet continues a previous row}
            \State $res_{agg}[1] \pluseq last\_packet\_output$
            \State $res_{agg}[0] \gets 0;\ rows_{fin}[0] \gets false$
        \EndIf
        % \State $rows_{fin} \gets find\_finished\_rows(x_{loc},\ x_{s\_old})$

        \LineComment{\textbf{4. Update Top-K values}}
        \For{$j \gets 0..B$} \Comment{Process only finished rows}
            \If{$res_{agg}[j] \geq worst_{curr}[j] \ \&\&\ rows_{fin}[j]$}
                \State $res_{loc}[j] \gets res_{agg}[j]$
                \State $res_{idx}[j] \gets row_{curr} + j - 1$
            \EndIf
            \State $argmin(res_{loc},\ worst_{curr}[j],\ worst_{idx}[j])$
        \EndFor
        \State $reset(res_{agg})$
        \EndFor
    \EndFunction
  \end{algorithmic}
\end{algorithm}

%% file: sections/experimental_results.tex
\section{Experimental Evaluation}\label{sec:experimental_results}

We employ a Xilinx Alveo U280 Accelerator Card equipped with 8 GB of
HBM2 memory (460 GB/s of total bandwidth over 32 channels) and a \texttt{xcu280-fsvh2892-2L-e} FPGA whose available resources are reported in \Cref{tab:resources}.
Our CPU baseline is \texttt{sparse\_dot\_topn}, a multi-threaded C++ implementation of Top-K \gls{spmv} \cite{sparsedottopn}, running on two Intel Xeon Gold 6248 and 384 GB of DRAM.
We also compare our design against GPUs, using a Tesla P100 (549 GB/s of HBM bandwidth): although we are not aware of any GPU implementation of \gls{topkspmv}, we can combine a fast GPU implementation of \gls{spmv} (such as cuSPARSE \cite{naumov2010cusparse}) with the GPU radix sort of Thrust \cite{bell2012thrust}, using both single and half-precision floating-point arithmetic. To provide a worst-case comparison, we also assume a zero-cost GPU sorting, as if cuSPARSE already retrieved Top-K values at no cost.
% While it is certainly feasible further to optimize \gls{topkspmv} on GPU, we can still derive precious insights about the memory efficiency and operational intensity of GPU \gls{spmv}, to understand if our FPGA design can offset the massive HBM bandwidth offered by modern GPUs and provide better overall performanc

We analyze 4 FPGA design: 32 bits unsigned fixed-point (\texttt{Q1.31}), 25 bits (\texttt{Q1.24}), 
20 bits (\texttt{Q1.19}), and a 32-bit floating-point version (\texttt{F32}) (\Cref{tab:resources}).
The number of \gls{hbm} channels limits the maximum number of cores to 32, although we could easily place more cores given our design's low resource footprint.
% For each bit-width, we use the highest number of cores giving a successful bitstream generation.
The CPU baseline uses floating-point arithmetic: our CPU does not support arbitrary reduced-precision, and simulated reduced-precision fixed-point resulted in lower performance. 
The experimental setup comprises 19 synthetic and real sparse embedding matrices (\Cref{tab:matrices_2}), with different sizes and distributions. Sizes are reported using \gls{bscsr} as in \Cref{fig:bscsr}: if stored as a na\"ive \gls{coo}, they would take 3 times as much space.
We simulate different non-zero distributions (uniform and left-skewed $\Gamma$), with 20 or 40 average non-zero entries per row (a sparsity factor of 2-8\%).
The use of synthetic matrices provides full control over the desired data-distribution, as the non-zero distribution in real embedding matrices is influenced by the chosen sparsification procedure. To the best of our knowledge, no public data-set of sparse embedding with sizes comparable to ours (millions of rows) is available. Instead, we sparsify the GloVe \cite{pennington2014glove} embedding corpus with the technique in \cite{mairal2009online}. Our performance is mostly unaffected by the underlying data-distribution (\Cref{sec:time}): as our FPGA design processes the matrix in a streaming fashion, density does not affect performance.
% Values are normalized in $L2$ norm to guarantee that similarities are scaled correctly and that we can use a single bit for the fixed-point integer part.
We perform each test 30 times, with different random vertices $\mathbf{x}$.

\input{tables/resources_2}
 
\input{tables/matrices_2}

\begin{figure}[t]
    \centering
    \includegraphics[width=\columnwidth]{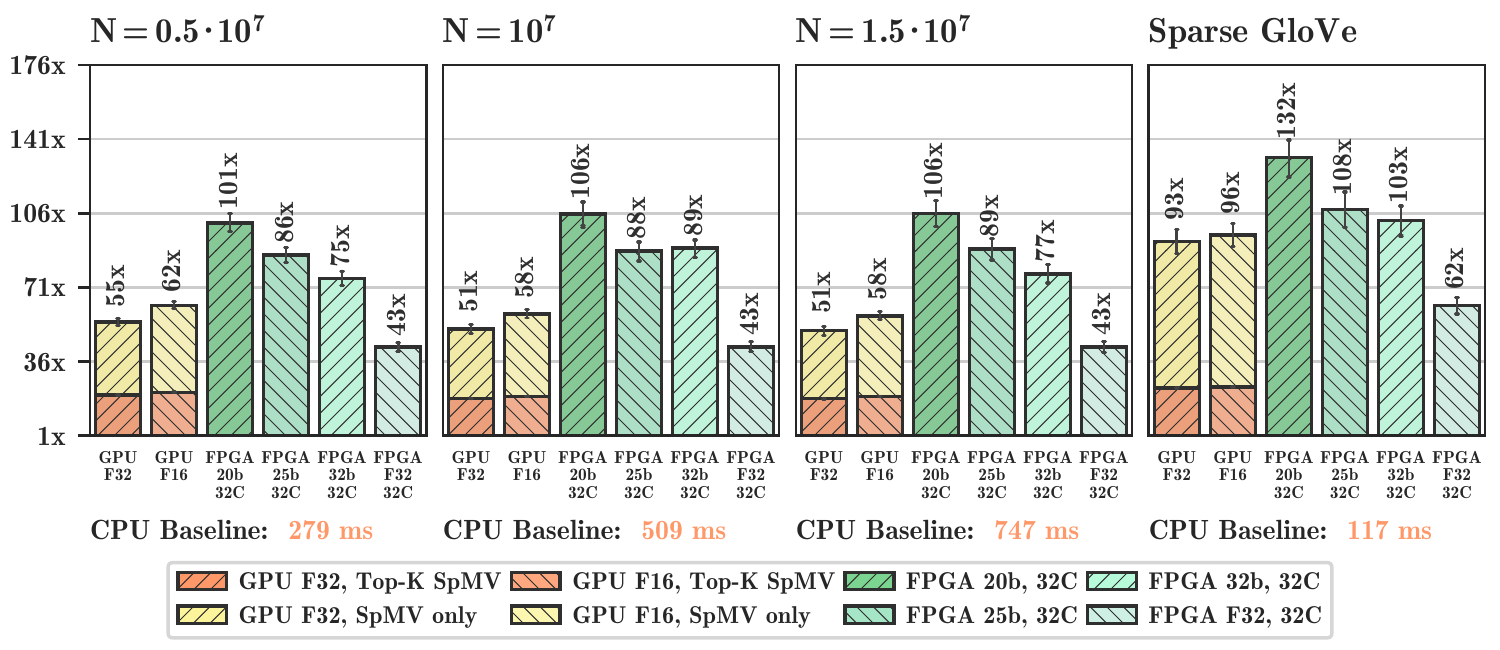}\\
    \caption{Execution time speedup on \gls{topkspmv} of our FPGA design and GPU vs. CPU. We can process over 57 billion non-zeros per second, 2x more than the GPU \gls{spmv}.}
    \label{fig:exec_times}
\end{figure}

\begin{figure}[t]
    \centering
    \includegraphics[width=\columnwidth]{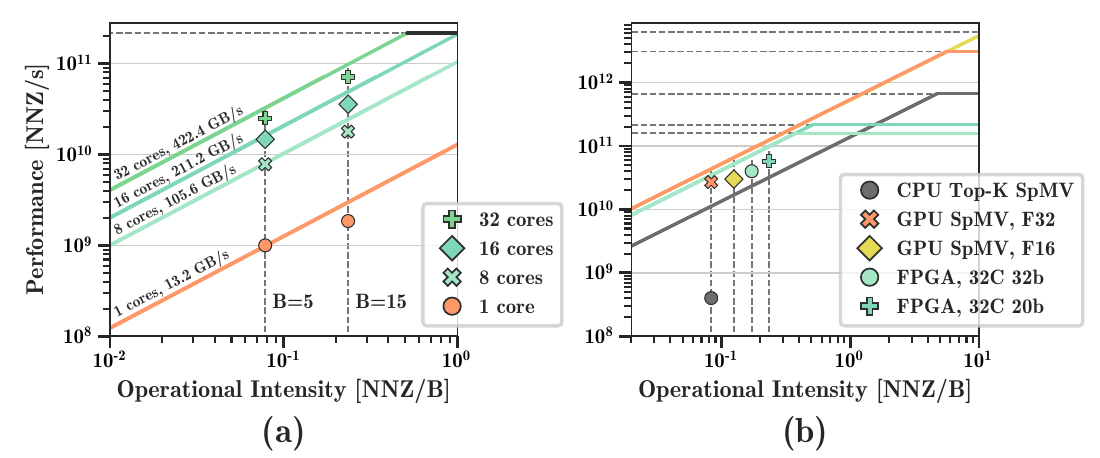}\\
    \caption{Roofline model of our \gls{topkspmv} architecture. \Circled{a} Operational intensity increase with \gls{bscsr}. \Circled{b} Comparison of FPGA vs. CPU and GPU. Thanks to \gls{bscsr}, we provide the highest operational intensity and the best performance.}
    \label{fig:roofline}
\end{figure}

\subsection{Execution Time}\label{sec:time}

For each design and matrix, we measure the speedup against CPU and GPU implementations and report results in \Cref{fig:exec_times}, for $K\!=\!100$.
While both GPU and FPGA implementations are significantly faster than the CPU baseline, our FPGA design achieves a 2x speedup over an idealized GPU \gls{topkspmv} implementation with zero-cost sorting, despite the 20\% lower peak memory bandwidth. When accounting for sorting costs on the GPU, this speedup can be as large as 7x. 
We expect to provide competitive performance even against a GPU with significantly higher memory bandwidth (e.g. Tesla A100, 1.5TB/s), and always provide greater power-efficiency (\Cref{sec:power}).
Fixed-point arithmetic guarantees higher speedups than floating-point, thanks to the pipelines' lower initiation interval.
Reduced-precision enables packing more non-zeros per transfer (higher $B$), resulting in increased operational intensity and better performance.
Our 32-cores design can find the Top-K values of a matrix with $10^7$ rows and 200 million non-zero entries in less than 4 ms and is suitable for real-time applications.
% Transfer time of $\mathbf{x}$ and of the results is included, and negligible w.r.t. the total time.

% Increasing the number of cores is always beneficial to performance, especially when processing large data-sets that mask scheduling costs and fixed costs.

\subsection{Power Efficiency}\label{sec:power}

Our FPGA hardware design consumes about 35W during execution, plus 40W for the host server, measured with an external power meter monitor.
% Changing bit-width did not significantly affect the power consumption.
The CPU implementation consumes around 300W during execution, while the GPU requires 250W (plus 40W for the host). Our fixed-point FPGA design provides 400x higher Performance/Watt ratio than the CPU, and 14.2x compared to the idealized GPU implementation (7.7x when accounting for an equal host machine): we provide higher performance without any sacrifice of power efficiency.

\subsection{Roofline Model Analysis}\label{sec:roofline}

From the Roofline Model in \Cref{fig:roofline}, built following the methodology in \cite{Siracusa2020RooflineFPGA}, we observe how the performance (expressed as non-zeros per second) of our FPGA design scales linearly to the total bandwidth of the available HBM channels: this result is significant as it entails predictable performance when deploying our design on an FPGA board with fewer (or more) HBM channels. Most importantly, \gls{bscsr} increases the operational intensity up to 3x compared to a na\"ive COO ($B\!=\!15$ vs. $B\!=\!5$), which immediately translates to an equivalent performance improvement. 
% On the right of \Cref{fig:roofline} we report performance against CPU and GPU. For the GPU, we provide a worst-case analysis by considering only the cost of \gls{spmv} and assuming zero-cost sorting. 
When compared against CPU and GPU, our FPGA design is the best in terms of operational intensity and bandwidth usage, both in absolute and percentage terms.
Despite having access to 20\% less bandwidth than the GPU and considering a \gls{topkspmv} GPU implementation with zero-cost sorting, we still provide the highest performance, thanks to the increased operational intensity of \gls{bscsr}.

\begin{figure}[t]
    \centering
    \includegraphics[width=\columnwidth]{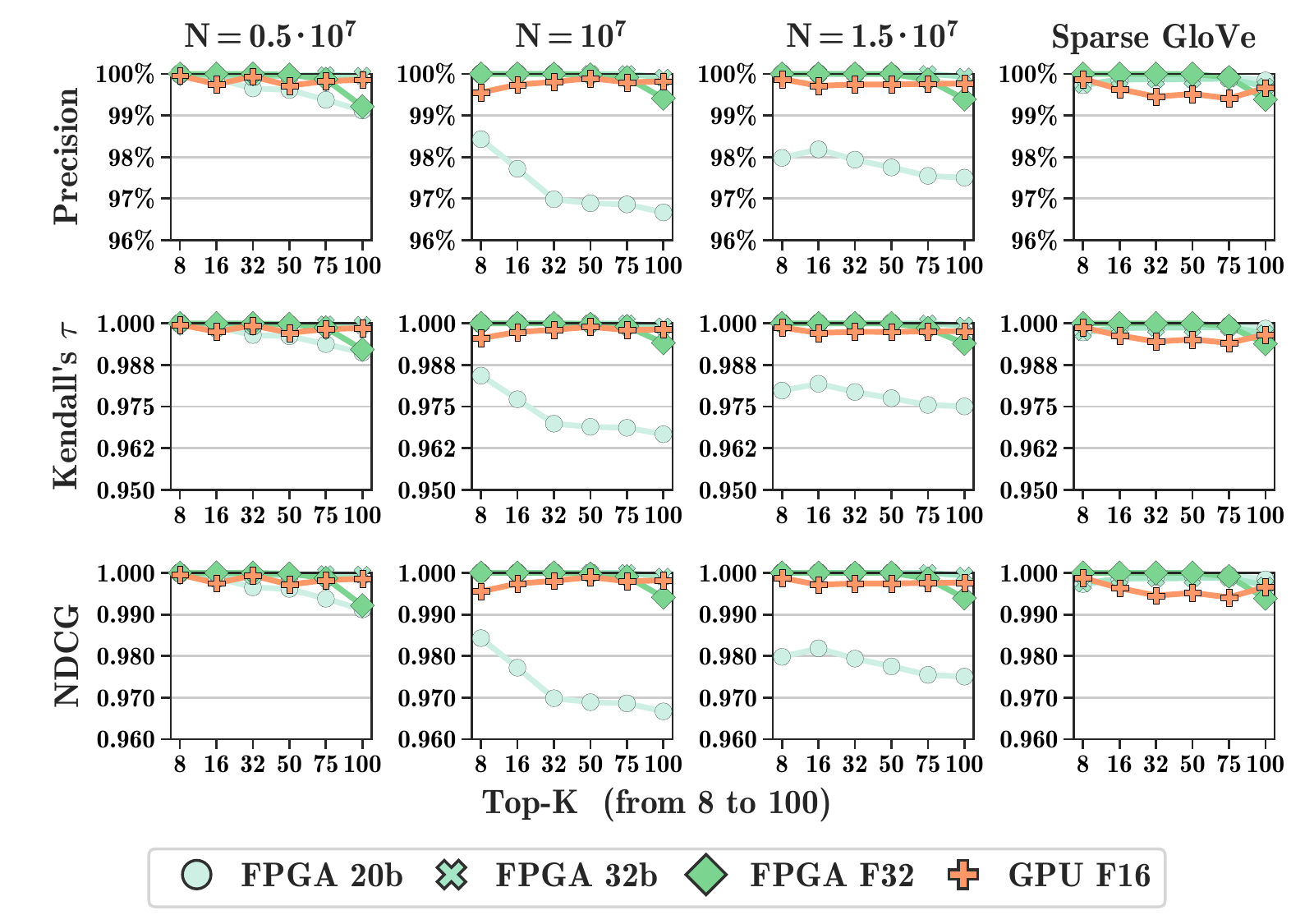}\\
    \caption{\gls{topkspmv} accuracy \textit{(higher is better)} for different architectures and types of reduced-precision arithmetic.}
    \label{fig:errors}
\end{figure}

%  we see that our FPGA design always reaches the maximum bandwidth available for a given number of HBM channels, and its performance scales almost linearly w.r.t. the number of cores and HBM channels. 
% To further improve performance, we need to increase the operational intensity, e.g. by processing more non-zero entries per clock cycle through data-compression.

% Besides improving performance, increasing the number is also beneficial to accuracy, as seen in \Cref{sec:accuracy}.

\subsection{Approximation Accuracy Analysis}\label{sec:accuracy}

We compare the accuracy of our approximated Top-K SpMV with the exact CPU results and with the approximated results of the GPU with half-precision floating-point. We look at values of K from 8 to 100 and evaluate accuracy using common Recommender System evaluation metrics such as Precision, Kendall's $\tau$, and NDCG \cite{shani2011evaluating}. Precision does not penalize out-of-order results; the other two metrics do. 
Our results are in line with the theoretical estimations in \Cref{tab:simulated_errors}, showing only a minor accuracy dip for large K. Even 20-bit fixed-point designs provide excellent accuracy across the board, with Precision above 97\% (\Cref{fig:errors}). Moreover, 32-bits fixed-point designs provide accuracy above the half-precision floating-point GPU implementation, despite our algorithmic approximation.

%% file: tables/resources_2.tex
\renewcommand\theadalign{tl}
\renewcommand\theadfont{\bfseries}
\setlength\tabcolsep{3pt}

\begin{table}
\centering
\ra{1.0}
    \caption{Resource usage, clock frequency and power consumption of our architecture. Best values in bold.
    % Floating point arithmetic greatly increases the use of DSPs and LUTs, while LUT utilization and clock frequency change linearly w.r.t bit-width.
    % Other bit-widths, omitted for brevity, show the same trends
    }
    \resizebox{1\linewidth}{!}{
	\begin{tabular}{@{}lllllllll@{}}
		\toprule
	% 2 Lines
% 	& \multirow{2}{*}{\thead{Bit-width}} & \multirow{2}{*}{\thead{BRAM}} & \multirow{2}{*}{\thead{DSP}} & \multirow{2}{*}{\thead{FF}} & \multirow{2}{*}{\thead{LUT}}  & \multirow{2}{*}{\thead{URAM}} &\multirow{2}{*}{\thead{Clock}} & \thead{Power\\Consumption}\\
    % 1 Line
	\multirow{2}{*}{\thead{Bit-width}} & \multirow{2}{*}{\thead{Cores}} & \multirow{2}{*}{\thead{LUT}} & \multirow{2}{*}{\thead{FF}} & \multirow{2}{*}{\thead{BRAM}} & \multirow{2}{*}{\thead{URAM}}  & \multirow{2}{*}{\thead{DSP}} & \thead{Clock\\(MHz)} & \thead{Power\\Cons.}\\
	   
	    \midrule
    	
		\textbf{20 bits} & 32 & 38\% & 35\% & 20\% & 33\% & \textbf{7\%} & \textbf{253} & \textbf{34 W} \\

% 		 \midrule
		\textbf{25 bits} & 32 & 38\% & 36\% & 20\% & 30\% & 11\% & 240 & 35 W \\

% 		\midrule
	    \textbf{32 bits} & 32 & \textbf{35\%} & \textbf{33\%} & 20\% & 27\% & 17\% & 249 & 35 W \\
	
    % 		\midrule
		\textbf{32 bits, float} & 32 & 44\% & 37\% & 20\% & \textbf{26\%} & 19\% & 204 & 45 W \\
        \midrule
        \textbf{Available} &  & 1097419 & 2180971 & 1812 & 960 & 9020 & & \\
		\bottomrule
	\end{tabular}
    }
    \label{tab:resources}
\end{table}

%% file: tables/matrices_2.tex
\renewcommand\theadalign{tl}
\renewcommand\theadfont{\bfseries}

\begin{table}
\centering
% \ra{1.2}
    \caption{Matrices in the evaluation, with $M = 512$ and $1024$, and with memory occupation using \gls{bscsr} as in \Cref{fig:bscsr}. \textit{Distribution} controls the number of non-zeros per row.}
    \resizebox{1\columnwidth}{!}{
	\begin{tabular}{@{}llll@{}}
		\toprule
		\thead{Distribution} & \thead{Rows}  & \thead{Non-zeros (min-max)} & \thead{Size (min-max, GB)} \\
		\midrule
	%	\textbf{$10^4$} & $249\,181$ & $24.91$     & $\SI{2.99}{\mega\byte}$ \\
		\multirow{3}{*}{Uniform} & $0.5 \cdot 10^7$ & $10^8 - 2\cdot 10^8$     & $\SI{0.4}{\giga\byte} - \SI{0.8}{\giga\byte}$ \\
                                 & $1.0 \cdot 10^7$ & $2\cdot 10^8 - 4\cdot 10^8$    & $\SI{0.8}{\giga\byte} - \SI{1.7}{\giga\byte}$ \\
                                 & $1.5 \cdot 10^7$ & $3\cdot 10^8 - 6\cdot 10^8$    & $\SI{1.2}{\giga\byte} - \SI{2.5}{\giga\byte}$ \\
        \midrule
		\multirow{3}{*}{$\Gamma(k=3, \theta = 4/3)$} & $0.5 \cdot 10^7$ & $9.7 \cdot 10^7 - 1.97\cdot 10^8$     & $\SI{0.4}{\giga\byte} - \SI{0.8}{\giga\byte}$ \\
                                 & $1.0 \cdot  10^7$ & $1.9\cdot 10^8 - 3.95\cdot 10^8$    & $\SI{0.8}{\giga\byte} - \SI{1.7}{\giga\byte}$ \\
                                 & $1.5 \cdot 10^7$ & $2.9\cdot 10^8 - 5.92\cdot 10^8$    & $\SI{1.2}{\giga\byte} - \SI{2.5}{\giga\byte}$ \\
        \midrule
        Sparsified GloVe & $0.2 \cdot 10^7$ & $2.4 \cdot 10^7 - 4.6\cdot 10^7$ & $\SI{0.1}{\giga\byte} - \SI{0.3}{\giga\byte}$ \\
		\bottomrule
	\end{tabular}
    }
    \label{tab:matrices_2}
\end{table}

%% file: sections/conclusion.tex
\section{Conclusion and future work}

\gls{topkspmv} is a cornerstone of \gls{ir} and recommender systems based on sparse embeddings similarity and must be computed guaranteeing real-time latencies and power efficiency. As these constraints are hardly met on general-purpose architectures, we presented a novel approximate FPGA multi-core design for \gls{topkspmv} that leverages HBM, fixed-point arithmetic, and our new \gls{bscsr} matrix layout to achieve high-performance and modularity over the FPGA resources.
Our 32-cores 20-bit FPGA design achieves 100x the performance of a state-of-the-art CPU baseline and 2x the GPU performance with 20\% higher bandwidth. We deliver real-time results on sparse matrices with hundreds of millions of values while keeping 14.2x higher power efficiency than the GPU.

Future work will focus on adaptive compressed matrix representations by reconfiguring the FPGA in terms of numerical precision to guarantee desired targets of accuracy or performance.
We will also apply our design to smaller FPGA accelerator cards: with similar memory bandwidth, the computation can be cheaper and even more power-efficient, with no performance loss.
% such as the Xilinx Alveo U50, which is equipped with HBM but offers fewer FPGA resources compared to the Alveo U280:
% given the flexibility of our multi-core design, we believe that even less powerful FPGAs would provide identical performance given the availability of similar memory bandwidth, making the overall computation cheaper and even more power-efficient.
% Finally, we will develop an optimized GPU implementation to test if our approximation strategy can translate to different hardware architectures.
Finally, we will study how to apply \gls{bscsr} to other computations and possibly other architectures.